\documentclass[prl,10pt,floatfix,amsmath,twocolumn,superscriptaddress]{revtex4-1}
\usepackage[ansinew]{inputenc}
\usepackage{graphicx}
\usepackage{pst-all}
\usepackage{color}
\usepackage{dcolumn}
\usepackage{amsmath}
\usepackage{amsthm}
\usepackage{bm}
\usepackage{layout}
\usepackage{float}
\usepackage{txfonts}
\usepackage{amsfonts}
\usepackage{amssymb}
\begin{document}
\title{Topological nodal line states and a potential
catalyst of hydrogen evolution in the TiSi family}

\author{Jiangxu Li$^{\ddagger,}$}
\affiliation{Shenyang National Laboratory for Materials Science,
Institute of Metal Research, Chinese Academy of Science, School of
Materials Science and Engineering, University of Science and
Technology of China, 110016 Shenyang, Liaoning, China.}

\author{Hui Ma$^{\ddagger,}$}
\affiliation{Shenyang National Laboratory for Materials Science,
Institute of Metal Research, Chinese Academy of Science, School of
Materials Science and Engineering, University of Science and
Technology of China, 110016 Shenyang, Liaoning, China.}
\affiliation{Environmental Corrosion Center, Institute of Metal
Research, Chinese Academy of Sciences, 110016, Shenyang, Liaoning,
China}

\author{Shaobo Feng}
\affiliation{Shenyang National Laboratory for Materials Science,
Institute of Metal Research, Chinese Academy of Science, School of
Materials Science and Engineering, University of Science and
Technology of China, 110016 Shenyang, Liaoning, China.}

\author{Sami Ullah}
\affiliation{Shenyang National Laboratory for Materials Science,
Institute of Metal Research, Chinese Academy of Science, School of
Materials Science and Engineering, University of Science and
Technology of China, 110016 Shenyang, Liaoning, China.}

\author{Ronghan Li}
\affiliation{Shenyang National Laboratory for Materials Science,
Institute of Metal Research, Chinese Academy of Science, School of
Materials Science and Engineering, University of Science and
Technology of China, 110016 Shenyang, Liaoning, China.}

\author{Junhua Dong}
\affiliation{Environmental Corrosion Center, Institute of Metal
Research, Chinese Academy of Sciences, 110016, Shenyang, Liaoning,
China}

\author{Dianzhong Li}
\affiliation{Shenyang National Laboratory for Materials Science,
Institute of Metal Research, Chinese Academy of Science, School of
Materials Science and Engineering, University of Science and
Technology of China, 110016 Shenyang, Liaoning, China.}

\author{Yiyi Li}
\affiliation{Shenyang National Laboratory for Materials Science,
Institute of Metal Research, Chinese Academy of Science, School of
Materials Science and Engineering, University of Science and
Technology of China, 110016 Shenyang, Liaoning, China.}

\author{Xing-Qiu Chen}
\email{xingqiu.chen@imr.ac.cn} \affiliation{Shenyang National
Laboratory for Materials Science, Institute of Metal Research,
Chinese Academy of Science, School of Materials Science and
Engineering, University of Science and Technology of China, 110016
Shenyang, Liaoning, China.}

\date{\today}

\begin{abstract}
{Topological nodal line (DNL) semimetals, formed by a closed loop of
the inverted bands in the bulk, result in the nearly flat
drumhead-like surface states with a high electronic density near the
Fermi level. The high catalytic active sites associated with the
high electronic densities, the good carrier mobility, and the proper
thermodynamic stabilities with $\Delta G_{H^*}$$\approx$0 are
currently the prerequisites to seek the alternative candidates to
precious platinum for catalyzing electrochemical hydrogen (HER)
production from water. Within this context, it is natural to
consider whether or not the DNLs are a good candidate for the HER
because its non-trivial surface states provide a robust platform to
activate possibly chemical reactions. Here, through first-principles
calculations we reported on a new DNL TiSi-type family with a closed
Dirac nodal line consisting of the linear band crossings in the
$k_y$ = 0 plane. The hydrogen adsorption on the (010) and (110)
surfaces yields the $\Delta G_{H^*}$ to be almost zero. The
topological charge carries have been revealed to participate in this
HER. The results are highlighting that TiSi not only is a promising
catalyst for the HER but also paves a new routine to design
topological quantum catalyst utilizing the topological DNL-induced
surface bands as active sites, rather than edge sites-, vacancy-,
dopant-, strain-, or heterostructure-created active sites.}
\end{abstract}


\maketitle

Topological semimetals\cite{TSM}, which have been classified into
topological Dirac semimetal
(TDs)\cite{TD,na3bi1,na3bi2,na3bi3,na3bi4,cdas1,cdas2,cdas3},
topological Weyl semimetals
(TWs)\cite{TW,taas1,taas2,taas3,taas4,taas5,ta3s2,tairte4,wte2,wte3,cts,
hgte,zrte}, and topological nodal line semimetals
(DNLs)\cite{Fang2016,TL-1,TL-2,TL-3,Li2016,dnl1,dnl2,dnl3,dnl4,dnl5,dnl6,dnl7}
and beyond \cite{TBS}, have currently attracting extensively
interest in condensed matter physics and materials science. In
difference from both TDs and TWs which exhibit isolated Dirac cones
and Weyl nodes in its bulk phase, the class of DNLs
\cite{Li2016,dnl1,dnl2,dnl3,dnl4,dnl5,dnl6,dnl7,dnl8,dnl9,dnl10}
shows a fully closed line nearly at the Fermi level in its bulk
phase. The projection of the DNLs into a certain surface would
result in a closed ring in which the topological surface states
(usually flat bands) occur due to the non-trivial topological
property of its bulk phase. This kind of exotic band structures
exhibit various novel properties, such as giant surface Friedel
oscillation in beryllium \cite{Li2016}, flat Landau level
\cite{Landau2015} and long-range Coulomb interaction
\cite{Moon2016}. Currently, only the DNL-induced topological surface
bands has been directly confirmed in beryllium \cite{Li2016} and the
DNLs have been, partially or indirectly, observed in several bulk
materials, such as PtSn$_4$ \cite{6},TlTaSe$_2$\cite{8} and
PbTaSe$_2$\cite{7} and ZrSiS\cite{9,10,11} as well as in a
two-dimensional DNL monolayer of Cu$_2$Si \cite{12}.

\begin{figure}[!h]
\centering
\includegraphics[width=0.48\textwidth]{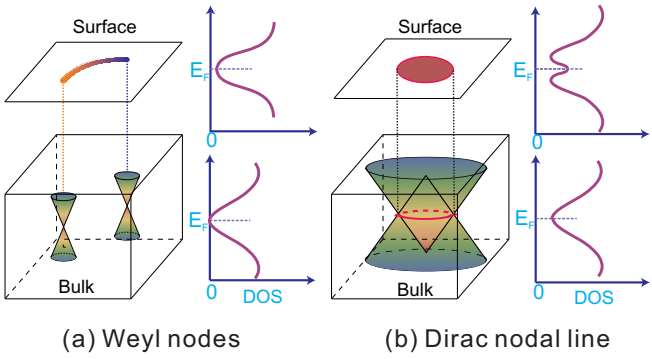}
\caption{Schematics of momentum space diagrams and density of states
(DOSs) of TWs and DNLs. Panel (a): A pair of Weyl nodes in bulk
(left lower panel) and broken Fermi arc surface states (left upper
panel) on the surface and their corresponding DOSs (right panels);
Panel (b) A DNL in bulk (left lower panel) and the nearly flat
drumhead-like non-trivial surface states on the surface (left upper
panel) and their corresponding DOSs (right panels).} \label{fig0}
\end{figure}

Most recently, TWs (NbP, TaP, NbAs and TaAs) have been considered to
serve as excellent candidates of catalysts because of the remarkable
performance of the hydrogen evolution reaction (HER)\cite{TWs-HRE9}.
This key concept of TWs as catalysts is extremely nice by
alternatively providing a way to create the active sites with
topological surface states, rather than by traditionally increasing
the active edge sites or vacancies
\cite{HER00,HER01,HRE3,HRE4,HRE5,HRE10,HRE11}. The possible
bottleneck of TWs as catalyst may be its much lower carrier density
around the Fermi level (Fig. \ref{fig0}a), because that the strength
of electrostatic screening in TWs is much weaker than the normal
metal (\emph{e.g.}, Pt). However, a DNL material shows two
distinguishing features from both TDs and TWs \cite{TWs-HRE9}. In
its bulk phase, a DNL results in a certain carrier density around
the Fermi level (Fig. \ref{fig0}b) and its topologically protected
nearly flat drumhead-like non-trivial surface states provide an
unusually high electronic density around the Fermi level (Fig.
\ref{fig0}b), as seen in pure metal beryllium \cite{Li2016}. Besides
these advantages, in similarity to both TDs and TWs the DNL-induced
surface states certainly provides sufficient active plane (Fig.
\ref{fig0}b) and the carrier mobilities are, in principle, high,
because the DNL is formed by the continuously linear crossings of
energy bands around the Fermi level (Fig. \ref{fig0}b). Therefore,
DNLs would fit better catalyst for the HER due to three combined
advantages: (\emph{i}) non-trivial drumhead-like surface states as
robust active sites, (\emph{ii}) good mobilities of carriers,
(\emph{iii}) a certain density of carriers around the Fermi level.
In addition, the crucial thermodynamic descriptor
\cite{HER00,HER01,HRE1,HRE3,HRE4,HRE5,HRE6,HRE7,HRE8,HRE10,HRE11} of
$\Delta G_{H^*}$ as good catalysts should be zero as close as
possible, which can be screened well through first-principles
calculations.

Within this context, through first-principles calculations (details
refer to Ref. \onlinecite{online}) we report a new DNL family of the
TiSi-type materials $MX$ ($M$ = Ti, Zr, Hf; $X$ = Si, Ge, Sn). The
DNL exists in the $k_y$ = 0 plane of the bulk Brillouin zone (BZ)
and induces the nearly flat drumhead-like topological non-trivial
surface states, thereby resulting in a highly high localized
electronic density around the Fermi level on the surface.
Interestingly, on the two (010) and (110) surfaces of TiSi the
hydrogen adsorption free energies $\Delta G_{H^*}$ are derived to be
almost zero, being much more closer to zero than those values of all
known catalysts for the HER\cite{HRE1,HRE3,HRE4,HRE5,HRE6,HRE7,HRE8}
including the most extensively used precious platinum (Pt).

\begin{figure}[!h]
\centering
\includegraphics[width=0.48\textwidth]{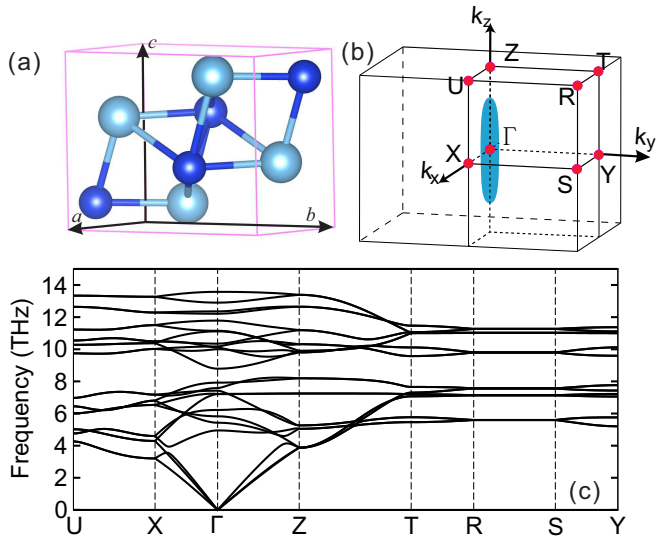}
\caption{Crystal structure, Brillouin zone (BZ) and phonon
dispersion of TiSi. Panel (a): the orthorhombic lattice with the
space group of $Pnma$, Panel (b): the BZ and high symmetrical
k-points of the lattice and the shaded region indicates the
corresponding position of the Dirac nodal line at the $k_Z$ = 0
plane, and Panel (c): the DFT-derived phonon dispersion.}
\label{fig1}
\end{figure}

The TiSi samples were first prepared by the arc melted method and
then annealed in vacuum for 48 hours at 1200 $^o$C. The X-ray
diffraction demonstrate that TiSi crystallizes in the orthorhombic
lattice (Fig. \ref{fig1}a) with the space group of \emph{Pnma} (No.
62) and the refinement reveals that Si occupies the Wyckoff 4$c$
site (0.0362, 0.2500, 0.1103) and Ti at another Wyckoff 4$c$ site
(0.1820, 0.2500, 0.6250). Our current experimental findings are
supported by our theoretical lattice constants $a$ = 6.529 \AA\,,
$b$ = 3.645 \AA\, and $c$ = 5.004 \AA\,, also in good accord with
the previous experimental data \cite{expt1, expt2,expt3,expt4}
(supplementary Table S1). In addition, the derived phonon dispersion
does not show any imaginary frequencies and is dynamically stable
(Fig. \ref{fig1}b).

\begin{figure}[hbt]
\centering
\includegraphics[width=0.48\textwidth]{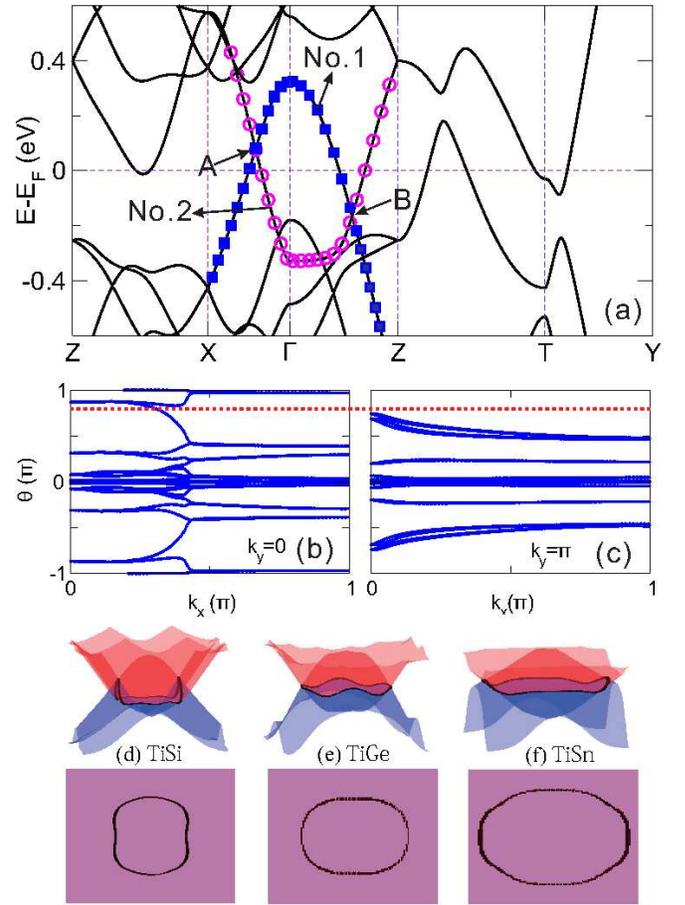}
\caption{Electronic band structure and the evolution of Wannier
centers of TiSi. Panel (a): Calculated electronic band structure
without the spin-orbit coupling (SOC) inclusion. In the panel (a),
the hollow circs denotes the weight of Ti $d_{yz}$-like states and
the solid squares show the weight of the Ti $d_{z^2}$-like states,
Panel (b and c): The evolution of Wannier centers along the $k_y$
direction. The evolution lines cross the reference line (dotted red
line) odd and even times in the $k_y$ = 0 and $\pi$ planes,
respectively. $k_y$ and $k_x$ are in the directions, as given in
Fig. 2b. Panels (d, e, and f): The Dirac nodal lines in the $k_y$ =
0 plane of TiSi, TiGe and TiSn, respectively. The upper and lower
panels denote their three-dimensional visualizations and their
corresponding two-dimensional projections on the (010) plane,
respectively.} \label{fig2}
\end{figure}

\begin{figure*}[hbt]
\centering \vspace{0.1cm}
\includegraphics[width=0.85\textwidth]{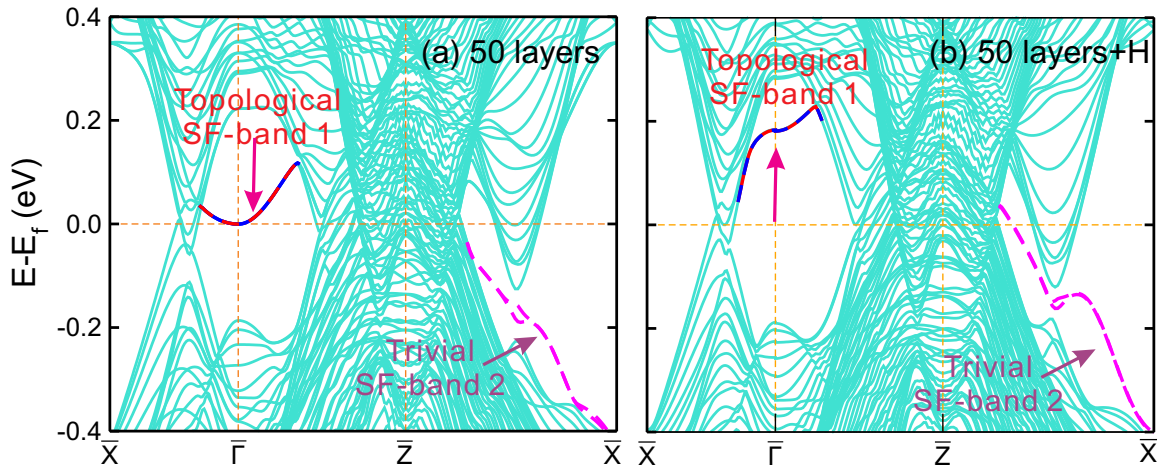}
\caption{The surface electronic band structures of the TiSi (010)
surface. Panels  (a and b): the comparison of the derived surface
electronic structures on the 50-atomic-layer (010) surface without
and with the top and bottom hydrogen adsorption, respectively. Noted
that the topologically protected non-trivial surface bands (marked
as the topological non-trivial SF-band 1) has been highlighted and
the trivial surface bands (trivial SF-band 2) has been also marked
in panels along $\bar{\Gamma}$-$\bar{X}$. }~\label{fig3}
\end{figure*}

In standard DFT calculations, as shown in Fig. \ref{fig2}a the bands
near the Fermi energy are mainly contributed from the Ti
\emph{d}-like orbitals. Without the SOC inclusion, there are the two
nearly linear band crossings, A and B points, as marked in Fig.
\ref{fig2}a. The one (A) locates at 0.1 eV above the Fermi level in
the X-$\Gamma$ direction and the other one (B) lies about 0.18 eV
below the Fermi level along the $\Gamma$-Z direction. They are
physically induced by the band inversion. At the centre of the
Brillouin zone (BZ, Fig. 1b), $\Gamma$, the $d_{yz}$ $\rightarrow$
$d_{z^2}$ band inversion (Fig. S1 \cite{online}) occurs between the
two bands No. 1 and No. 2, as marked in Fig. \ref{fig2}a.
Strikingly, the band crossings not only appear at these two points,
but also form a circle-like closed line around the $\Gamma$ point in
the k$_y$=0 plane (Fig. \ref{fig2}d). This is the apparent sign of
the DNL appearance. The band crossings between No.1 and No.2 bands
do not occur at the same energy level, but show a wave-like closed
curve upon the $k$ vectors around the centered $\Gamma$ point.
Certainly, this DNL stability is highly robust, protected by the
inversion and time-reversal symmetry without the spin-orbit coupling
(SOC) effect. Because of the light masses of Ti and Si, its SOC
effect is rather weak; therefore, it does not affect the electronic
band structure, apparently (Fig. S2\cite{online}). Furthermore, the
non-trivial topology order of TiSi is confirmed by the non-Abelian
Berry connection method \cite{Waner2011,Waner20112,Sun2016}, as
shown in Fig. \ref{fig2}(b) and \ref{fig2}(c). In the $k_y$ = 0
plane in which the DNL locates, the loop of the Wannier center
evolution change partners from $k_x$ = 0 to $k_x$ = $\pi$. However,
In the $k_y$ = $\pi$ plane, no partner changes. Hence, the evolution
loop of the Wannier center cuts the reference line odd times in the
$k_y$ = 0 plane, whereas the crossing between Wannier center
evolution loop and the reference line zero or even times in the
$k_y$ = $\pi$ plane.

\begin{figure*}[hbt]
\centering \vspace{0.1cm}
\includegraphics[width=0.85\textwidth]{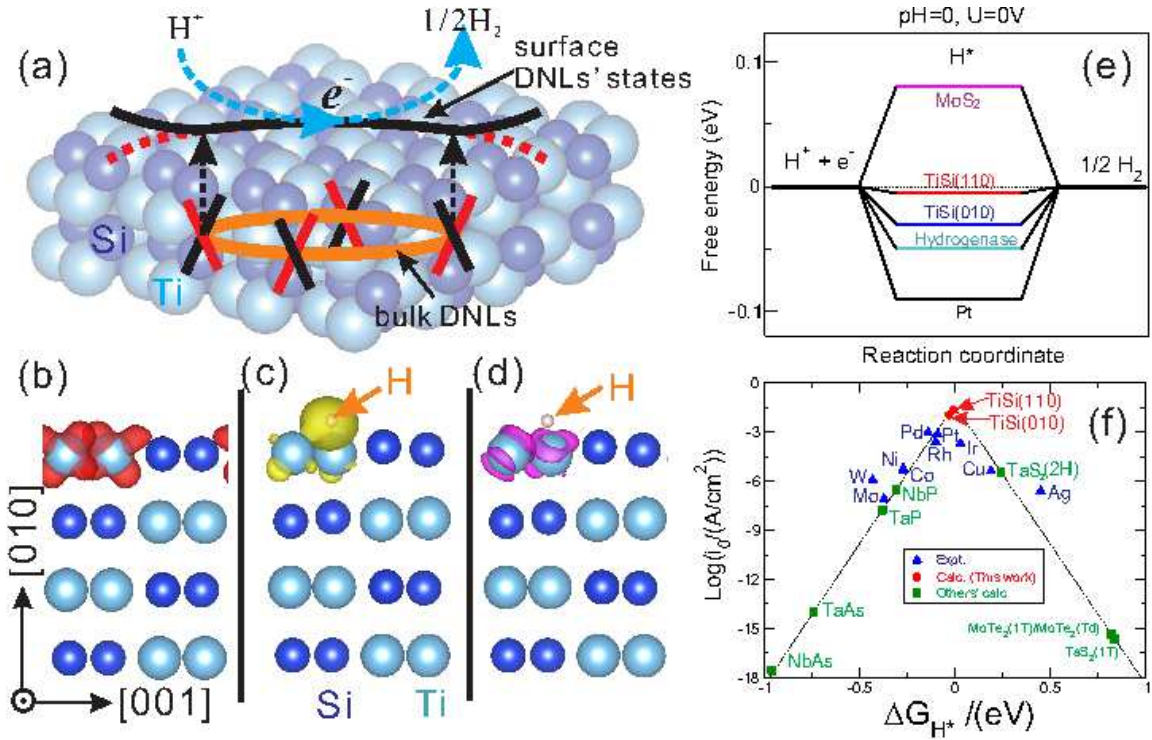}
\caption{The effects of the topological DNLs on the HER activity of
TiSi. Panel (a): Schematic of the HER reaction of the DNL-induced
half-filled nontrivial surface states to provide active plane and
the DNL states to provide high mobile and quickly diffused
electrons. Panel (b): the visualized localized charges of the
topological states on the (010) surface. The charges are
characteristic of $d_{yz}$-like orbitals from the topmost Ti atoms.
Panel (c): The visualized localized charge accumulations of hydrogen
on the (010) surface. Panel (d): the visualized localized charge
depletion surrounding Ti atoms. Panel (e): Free energy versus the
reaction coordinate of the HER of TiSi in comparison with several
other compounds. The data of MoS$_2$ (-0.08 eV for the edge states),
Pt (-0.09 eV) and Hydrogenase (-0.05 eV) are all taken from Ref.
\cite{HER01}. Panel (f): Volcano plot for the HER of TiSi in
comparison with various pure metals (the experimental data
\cite{HRE8} of Pt, Pd, Ni, Ir, Co, Rh, Ag, Cu, Mo and W), TWs (the
calculated data \cite{TWs-HRE9} of NbP, TaP, NbAs and TaAs), and
other candidates (the theoretical data \cite{TWs-HRE9} of
TaS$_2$(2H), MoTe$_2$(1T$^\prime$), MoTe$_2$(Td), and TaS$_2$(1T)).
}~\label{fig4}
\end{figure*}

We have also considered the isoelectronic and isostructural TiGe,
TiSn, HfSi, HfGe, HfSn and ZrSi, ZrGe and ZrSn. As shown in
supplementary Table S1 and Fig. S1-S6 \cite{online}, the electronic
band structures of ZrSi and GeSi are qualitatively the same physics
as TiSi does (see Fig. \ref{fig2}(e and f)).

To inspect the topological surface bands for TiSi, we have
calculated the electronic structures of the (010) surface by varying
the thickness of the slabs (Fig. S7\cite{online}). As expected, the
robust surface electronic bands (topological SF-band 1 in Fig.
\ref{fig3}a) appear, when the slab's thickness is above eight atomic
layers along the \emph{b}-axis. From Fig. \ref{fig3}a, outside the
Dirac nodal ring on the (010) surface projected by the DNL of its
bulk phase, the two-fold degenerated topological SF-band 1 clearly
separates: one goes to the unoccupied conduction bands integrating
with the projection of the electronic bands of bulk phase and the
other one emerges into the valence bands overlapping with the
projected bulk bands. With other words, these separated surface
bands outside the projected Dirac nodal ring are the topologically
trivial states and not correlated with the bulk DNL states. These
separated trivial surface bands mainly originates from the Ti
\emph{$d_{xz}$}-, \emph{$d_{xy}$}- and $d_{x^2-y^2}$-like states.
This means that the topological non-trivial surface bands SF-band 1
only occur within the DNL-projected Dirac nodal ring on the (010)
surface. The topologically protected SF-band 1 around the
$\bar{\Gamma}$ point are mainly comprised with the $d_{yz}$ and
$d_{z^2}$-like electronic states from the topmost atomic layer,
reflecting well the $d_{yz}$ $\rightarrow$ $d_{z^2}$ band inversion
in its bulk phase (Fig. 2a). This SF-band 1 is two-fold degenerated,
half-filled when the surface is electrically neutral, in similarity
to the case of Be \cite{Li2016}.

Importantly, the three main features of the DNLs in TiSi motivates
us to consider its activities as catalysts, as conceptually shown in
Fig. \ref{fig4}a. Firstly, the nearly flat drumhead-like non-trivial
topological surface states (SF-band 1 in Fig. \ref{fig3}a) on the
(010) surface disperses parabolically and its lowest-energy part
exactly cuts the Fermi level of 0 eV at $\bar{\Gamma}$, suggesting
the possibility of robust active planes for catalysis against
defects, impurities, and other surface modifications. Secondly, from
Fig. \ref{fig2}a the Dirac nodal points on the NDLs around the Fermi
level are expected to exhibit high mobility because the linear band
crossing. In similarity to TDs and TWs, it will be favorable for the
free and quick diffusion of electrons. Thirdly, the topology carrier
density is not low due to the DNL presence around the Fermi level.

Following the theoretical suggestions
\cite{HER00,HER01,HRE1,HRE3,HRE4,HRE5,HRE6,HRE7,HRE8,HRE10,HRE11},
we evaluate the HER activities of the two (010) and (110) planes,
where the hydrogen adsorption free energy $\Delta G_{H^*}$ was
determined by varying different adsorption sites on the specified
surface (see method\cite{online}). Theoretically, $\Delta G_{H^*}$
is known to scale with activation energies and has been successfully
used as a descriptor for correlating theoretical predictions with
experimental measurements of catalytic activity for various
systems\cite{HER00,HER01,HRE1,HRE3,HRE4,HRE5,HRE6,HRE7,HRE8,HRE10,HRE11}.
The previous theoretical suggestions that for the best activity the
optimal value of $\Delta G_{H^*}$ should be 0 eV, where hydrogen is
bound neither too strongly nor weakly with active sites on the
surface \cite{HRE1}. Surprisingly, our calculated results
demonstrate that $\Delta G_{H^*}$ of HER on different TiSi surfaces
are very close to zero. It shows us that on the (010) surface
$\Delta G_{H^*}$ = -0.03 eV when hydrogen bridges two nearest
neighboring Ti atoms (Fig. \ref{fig4}c) and -0.05 eV with hydrogen
bridging Ti and Si on the topmost atomic layer. The (110) surface
even yields, $\Delta G_{H^*}$ = -0.005 eV, of an almost zero value.
In comparison with some typical catalysts (MoS$_2$, Pt and
hydrogenase) in Fig. \ref{fig4}e, the HER activities on the TiSi
surfaces are highly attractive. It can be seen that the values of
$\Delta G_{H^*}$ of TiSi show a much closer value to zero than both
the typical catalysts of Pt ($\Delta G_{H^*}$ = -0.09
eV)\cite{HER01} and the edge states ($\Delta G_{H^*}$ = 0.082
eV)\cite{HER01} of MoS$_2$. Furthermore, we have plot the Volcano
curves for the HER of TiSi in comparison with some data known.
Remarkably, among all known data TiSi exhibits a $\Delta G_{H^*}$
most close to zero. In particular, in comparison with these
TWs\cite{TWs-HRE9} in Fig. \ref{fig4}e, TiSi possibly shows a more
excellent HER performance, because their $\Delta G_{H^*}$ values of
TiSi are almost at the top of the Volcano curve. However, the
$\Delta G_{H^*}$ of both NbP and TaP are much lower than that of Pt,
and their asenides even have the corresponding values as negative as
-0.75 eV and -1.0 eV, respectively.

Mechanically, the calculations demonstrate that, after the hydrogen
adsorption on the (010) surface, the topological SF-band 1 (Fig.
\ref{fig3}b) becomes unoccupied above the Fermi level. The hydrogen
atom will obtain the charges and disperse in the deep energy region
below the Fermi level. This process can be made more clear by
visualizing the local charges in Fig. \ref{fig4}(b,c and d). On the
clean (010) surface, the charges of the topological SF-band 1 are
clearly localized at two nearest neighboring Ti atoms with the
$d_{yz}$ orbitals (Fig. \ref{fig4}b). After the hydrogen adsorption,
the topological charges are indeed transferred to the hydrogen. As
evidenced in Fig. \ref{fig4}c, a lone-pair \emph{s}-like orbital
appears in terms of the charge accumulations. Correspondingly, the
charge depletion of the two nearest neighboring Ti atoms are highly
visualized in Fig. \ref{fig4}d, which also refers to the position of
the localized topological charges on the H-free adsorption (010)
surface in Fig. \ref{fig4}b. This fact indicates that the topology
carrier on the SF-band 1 states are fully transferred into hydrogen
$s$-like orbitals, indicating that the bulk DNLs play an important
role in the HER as a potential catalyst.

Summarizing, we have reported on the new DNLs' family (TiSi, TiGe
and ZrSi) and theoretically demonstrate that they have the promising
potential as excellent catalyst for the HER performance because of
the active sites provided by the robust nearly flat drumhead-like
non-trivial surface states, a stable supply of itinerant electrons
from the certain carrier density and the high mobilities related
with the DNLs, and the most suitable $\Delta G_{H^*} \approx$ 0 for
the HER.

\bigskip
\noindent {\bf Acknowledgments} Work was supported by the ``Hundred
Talents Project'' of the Chinese Academy of Sciences and by the
National Natural Science Foundation of China (Grant Nos. 51671193
and 51474202) and by the Science Challenging Project No.
JCKY2016212A504. All calculations have been performed on the
high-performance computational cluster in the Shenyang National
University Science and Technology Park and the National
Supercomputing Center in Guangzhou (TH-2 system)


\bigskip
\noindent $^{\ddagger}$ These authors contributed equally to this work.\\

\end{document}